\documentclass[twocolumn,showpacs,preprintnumbers,amsmath,amssymb]{revtex4}
\usepackage{graphicx}
\usepackage{dcolumn}
\usepackage{bm}

\begin{document}

\title{Breakdown of the Migdal approximation at Lifshitz transitions with giant zero-point motion in the H$_3$S superconductor}

\author{Thomas Jarlborg$^{1,2}$, Antonio Bianconi$^{2,3,4}$ }

\affiliation{
$^1$ 
DPMC, University of Geneva, 24 Quai Ernest-Ansermet, CH-1211 Geneva 4, Switzerland
\\
$^2$ 
RICMASS, Rome International Center for Materials Science Superstripes, Via dei Sabelli 119A, 00185 Rome, Italy
\\
$^3$
Solid State and Nanosystems Physics, 
National Research Nuclear University MEPhI (Moscow Engineering Physics Institute),  Kashirskoye sh. 31, Moscow 115409, Russia,
\\
$^4$
Institute of Crystallography, Consiglio Nazionale delle Ricerche, via Salaria, 00015 Monterotondo, Italy}


\begin{abstract}
While 203 K high temperature superconductivity in H$_3$S has been interpreted by BCS theory in the dirty limit here we focus on the effects of hydrogen zero-point-motion and the multiband electronic structure relevant for multigap superconductivity near Lifshitz transitions. 
We describe how the topology of the Fermi surfaces evolves with pressure giving different Lifshitz-transitions. A neck-disrupting Lifshitz-transition (type 2) occurs 
where the van Hove singularity, vHs, crosses the chemical potential at 210 GPa and new small 2D Fermi surface portions appear with slow Fermi velocity 
where the Migdal-approximation becomes questionable. We show that the neglected hydrogen zero-point motion ZPM, plays a key role at Lifshitz transitions.
 It induces an energy shift of about 600 meV of the vHs.  The other Lifshitz-transition (of type 1) for the  appearing of a new Fermi surface 
 occurs at $130 GPa$ where new Fermi surfaces appear at the $\Gamma$ point of the Brillouin zone here 
 the Migdal-approximation breaks down and the zero-point-motion induces large fluctuations. 
 The maximum $T_c=203 K$ occurs at 160 GPa where $E_F/\omega_0=1$ in the small Fermi surface pocket
 at $\Gamma$.  A Feshbach-like resonance between a possible BEC-BCS condensate at $\Gamma$  
 and the BCS condensate in different k-space spots is proposed.
\end{abstract}

\pacs{74.20.Pq,74.72.-h,74.25.Jb}

$How$   $to$  $cite$  $this$   $article:$ \footnote{ How to cite this article:  Jarlborg, T. and Bianconi, A. Breakdown of the Migdal approximation at Lifshitz
transitions with giant zero-point motion in $H_3S$ superconductor. Sci. Rep. 6, 24816; doi: 10.1038/srep24816 (2016)}

\maketitle
 {\bf Introduction}
\\

The recent discovery of superconductivity in pressurized sulfur hydride metal with critical temperature  
with T$_c$ above 200K \cite{droz2,ere0} has provided experimental evidence that the  
coherent quantum macroscopic superconducting phase 
can occur at a temperature higher than 
the lowest temperature ever recorded on Earth $($-89.2 $^o$C)  \cite{bia1,bia2}.
High temperature superconductivity has been found in other hydrides \cite{ere2} like
pressurized $PH_3$ with T$_c$ at about 100 K \cite{droz3}, 
and it has been predicted in other hydrides like yttrium hydride \cite{yLi1}.
Superconductivity in pressurized hydrides was proposed by Ashchroft and his collaborators
\cite{ash4,ash5}. The Universal Structure Predictor: Evolutionary Xtallography, USPEX, code \cite{zur} available today has allowed 
to predict the structure and high temperature superconductivity in pressurized sulfur hydride 
 \cite{yLi,duan,duan2}.  
The prediction of metallic H$_3$S with the $Im\bar{3}m$ lattice symmetry  \cite{duan,duan2} 
has now been confirmed by x-ray diffraction experiments \cite{ere0} above 120 GPa while at lower pressure different stoichiometry and structure are found \cite{yLi2,ishi}.
The low mass of H atom is pushing the sulfur-hydrogen $T_2$$_u$ stretching mode and the $T_1$$_u$ phonons at $\Gamma$ to high energy, 
so the energy cut-off for the pairing interaction $\omega_0$ =150 $\pm50$ meV . 
The superconducting temperature was predicted
by employing the Allen-Dynes modified McMillan formula \cite{yLi,duan,duan2},
and the Migdal Eliashberg  formula 
 \cite{papa,errea,dura,qpic}.
Most of these works have used the Eliashberg theory with BCS approximations for isotropic pairing in a single band metal,
assuming the dirty limit, which reduces  a multiband to an effective single band metal, and  the Migdal approximation $ \omega_0/E_F <<1$.
More advanced theories have used  density-functional theory including calculations of the effective Coulomb repulsion \cite{sanna,ari}. 
Within the Eliashberg theory the Migdal approximation assumes that
the electronic and ionic degrees of freedom
can be rigorously separated in agreement with the Born-Oppenheimer approximation, 
which is valid for metals where the chemical potential is far away from the band edges. 
The breakdown of Migdal approximation was observed
 in the multi-gap superconductor $Al_xMg_{1-x}B_2$
\cite{mgb1,mgb5,boeri,mgb7} where the band edge of the $\sigma$ band fluctuates across the chemical potential due to zero point motion \cite{mgb4}
The breakdown of the Migdal approximation in a multigap superconductor is relevant since it requires 
 the correction to the chemical potential induced by pairing below the critical temperature
 \cite{mgb7} ignored in the standard Eliashberg theory. Moreover where the Migdal approximation breaks down in a multigap superconductor  it enetrs in a unconventional 
 superconducting phase where the coexisting multiple condensates 
could be either in the Bose Einstein condensate (BEC) regime or in the BEC-BCS crossover regime.
In these complex multi gaps superconductors where BEC, BEC-BCS and BCS condensates coexist
 the exchange interaction between
 different condensates  \cite{mgb7,shape,shape2,Guidini} which is neglected in the BCS approximations,
 but could become a relevant term both to increase or to stabilize high temperature superconductivity.
 This quantum term is a contact interaction, given by the quantum
 overlap between the condensates, which increases 
 the condensation energy \cite{mgb7,shape,shape2,Guidini,iron1,iron3} and the critical temperature
via the shape resonance, analog to the Fano Feshbach resonance in ultracold gases.
It has been found that the critical temperature shows the maximum amplification where one of the condensates is 
in the BCS-BEC crossover, which can occurs on the verge
of a Lifshitz transition \cite{bia1,bia2,iron1,iron3,bia3,bia8,bia4,bia5} 
At the Lifshitz transition, a change of the topology of the Fermi
surface is induced by pressure or doping and it has been shown to control
high temperature superconductivity in iron pnictides \cite{iron1,iron3}. 

Since $H_3$S is a multiband metal and Lifshitz transitions could occur by increasing pressure \cite{bia1,bia2}
it has been proposed that it is a multigap superconductior near Lifshitz transitions \cite{shape,shape2,Guidini}, 
where also multi scale phase separation  \cite{bia3} at a the Lifshitz transition could appear,
similar to what has been observed in the cuprates \cite{bia4,bia5,poc,campi}.

The coupling between the electronic and the 
atomic lattice degrees of freedom in sulfur hydrides at zero temperature
has been neglected in previous calculations of the electronic structure assuming a very large Fermi energy. 
On the contrary in the case of band edges close to the chemical potential and near Lifshitz electronic topological transitions 
the zero point motion (ZPM) cannot be neglected.  It is known that the ZPM modifies
the band structure itself \cite{capaz,gomez,can,wilk,fesi,bron,cevib}. 
with corrections of the band gap energy which can be  
larger than those induced by correlation. Moreover, the lattice disorder from ZPM will also cause band broadenings,
as it has been demonstrated in early investigations \cite{wilk}.
Such effects have been shown to be important for many properties in several different materials, 
even if their atomic masses are larger \cite{fesi,bron,cevib} and the lattice disorder can perturb spin waves and phonons \cite{eri14}.
The zero point motion is expected to be large in H$_3$S because of the small mass of H atoms, high frequency S-H stretching modes, 
and the double well potential for hydrogen in H$_3$S \cite{bia1} which is similar to the well studied cases of ice and biological macromolecules.

In this work we discuss the effect of lattice compression and zero point motion on the electronic structure of 
H$_3$S in the pressure range above 120 GPa where the metallic $Im\bar{3}m$ phase is stabilized by the ZPM of hydrogen atoms.
We  first  focus on the effect of the lattice compression on the large Fermi surface, identified with No.2 in ref. \cite{bia1} and 
on the small hole Fermi surface pockets at $\Gamma$ identified with No. 4 and No.5 in reference \cite{bia1}.
In particular we discuss the character of van Hove singularity (vHs) in the large Fermi surface, giving a peak in the density of states, 
and how its moves toward the chemical potential  
by increasing pressure. Above the critical pressure 210 GPa the vHs crosses 
the chemical potential and we find that tubular Fermi surface portions with two-dimensional character in the k-space appear, 
indicating a 3D-2D topological Lifshitz transition of type 2.
Second, we study the effect of the zero point motion on the van Hove singularity and we find a very 
large energy shift of the order of 600 meV of the vHs due to ZPM.
Third in the pressure range 120-180 GPa the amplitude of ZPM of hydrogen-sulfur stretching mode is shown to be 
larger than the S-H bond splitting in the R3m structure 
 therefore it stabilizes the symmetric $Im\bar{3}m$ structure. 
In this pressure range we have studied the Lifshitz transitions of type 1,  
for the appearing of new small Fermi surface pocket at $\Gamma$  at 130 GPa.
Finally we compare our results with recent experiments showing that
the isotope coefficient for the superconducting critical temperature diverges 
and the critical temperature goes toward zero  at 130 GPa in agreement with 
predictions of the BPV  theory \cite{shape,shape2,Guidini,iron1,iron3} for a Lifshitz transition of type 1.
Moreover the maximum critical temperature, 203 K,
appears where the Fermi energy in the small Fermi hole pocket at  $\Gamma$  is of the order
of the pairing interaction  $\omega_0$ as predicted in ref. \cite{shape,shape2,Guidini,iron1,iron3}.
\\
\\ 
\maketitle
 {\bf The Band Structure}
\\

The high pressure phase of metallic H$_3$S  has the cubic Space Group:  229  with $Im\bar{3}m$ lattice symmetry. 
The $Im\bar{3}m$  lattice structure can be described by the small Body Centered Cubic  (bcc) unit cell,
which has been used  by all previous calculations providing the electronic band dispersion in the bcc Brillouin zone (bcc BZ). 

Here, we use also an alternative simple cubic unit cell made by 8 atoms
per unit cell with a simple cubic  Brillouin zone (sc BZ),
which permits an easier comparison with a traditional group 
of superconductors, namely the A15 compounds. In fact A15 
compounds have a lattice structure with the cubic Space Group: 223 
belonging to the same ditesseral central class, or galena type,
of cubic space groups with the same Hermann-Mauguin point group $m\bar{3}m$.

\begin{figure}
\includegraphics[height=8.0cm,width=8.0cm]{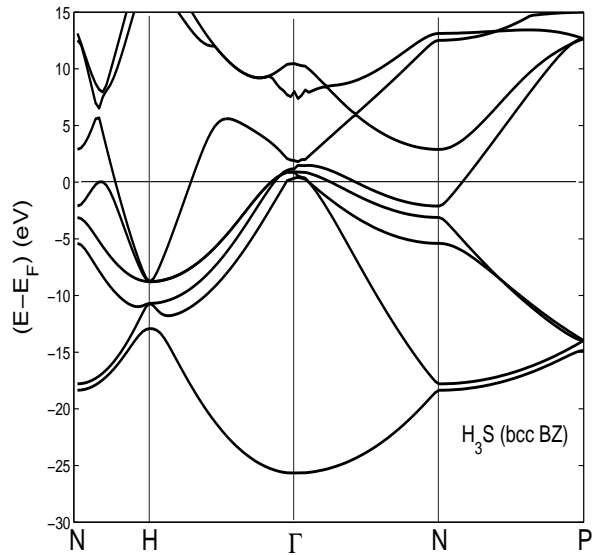}
\caption{(Color online) 
The band structure for H$_3$S at symmetry points 
for the small bcc unit cell with $a$=5.6 a.u., corresponding to $P=210 GPa$ in the bcc Brillouin zone (bcc BZ)}
\label{fig3bcc}
\end{figure}

\begin{figure}
\includegraphics[height=8.0cm,width=8.0cm]{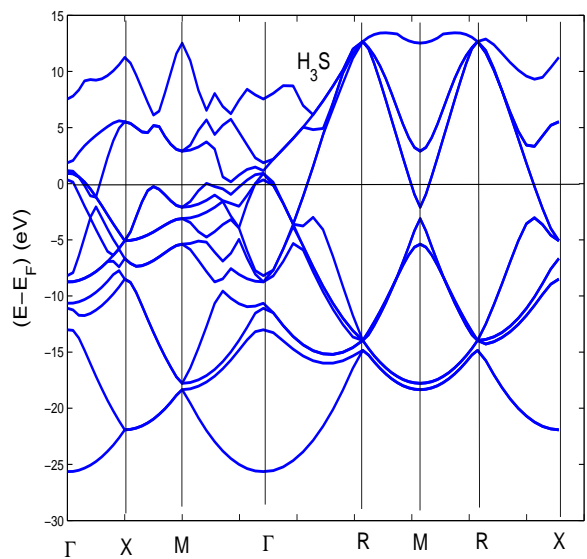}
\caption{(Color online) 
The band structure for H$_3$S at symmetry points for the simple cubic large unit cell, in the simple cubic BZ at a pressure of $210 GPa$}
\label{fig4bcc}
\end{figure}

\begin{figure}
\includegraphics[height=8.0cm,width=8.0cm]{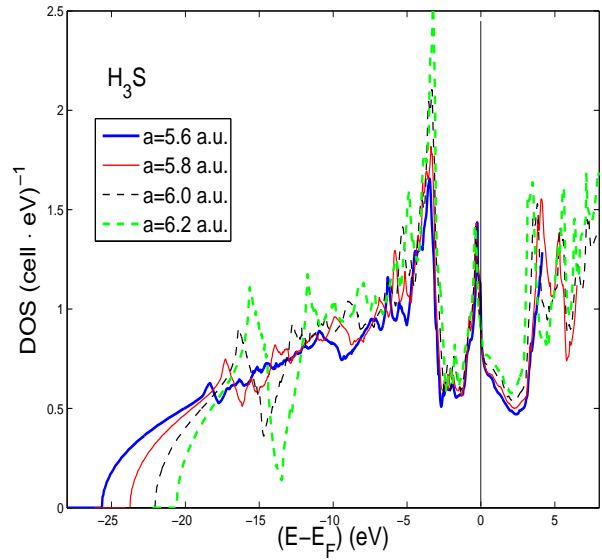}
\caption{(Color online) The total DOS for H$_3$S with variable lattice constants
between 5.6 and 6.2 a.u. }
\label{fig1}
\end{figure}

We have obtained the band dispersion in the large simple cubic
 (sc) Brillouin zone which grabs more details of the electronic band 
dispersion in the complex bitruncated cubic honeycomb lattice.
Here S sites form a bcc lattice (exactly as Si in the A15 compound V$_3$Si) with linear chains of H on the limits of the sc
cell, similar to how the transition metals form linear chains in A15 superconductors. These results 
allow a more easy  comparison of the electronic bands between the two type of materials.
 
We show the electronic self-consistent paramagnetic bands for the bcc BZ in Fig. 1 
and in Fig. 2 for the sc BZ.  In this last picture the simple cubic unit cell contains 8 sites totally.
The calculations have been performed by the linear muffin-tin orbital (LMTO) method \cite{bdj} and the
local spin-density approximation (LSDA) in the frame of standard methods \cite{jmp,ce,mgj,jb}. 
The basis set goes up through $\ell$=2 for S and $\ell=1$ for H. 
We have found that the S-2p core levels are 
always far below the valence band region at these pressures (about 12.2 Ry below $E_F$).
The Wigner-Seitz (WS) radii are 0.38$a$ for S
and 0.278$a$ for H.
The k-point mesh
corresponds to 11 points between $\Gamma$ and $X$, or 1331 points totally in 1/8 of the BZ,
or with finer k-point mesh for plots of the bands along symmetry lines.
One spin-polarized calculation is made for the largest volume starting from an imposed ferro-magnetic (FM) configuration.
All local FM moments converge to zero, which shows that FM and FM  spin fluctuations are unlikely.
The low-lying s-band on S is very similar to
the Si-s band in the A15-compound V$_3$Si \cite{jmp}, but the high pressure 
in H$_3$S makes it to overlap
with the S-p band. In contrast, the Si-p band in V$_3$Si is separated from the Si-s band. 
The separation is never complete in H$_3S$, but it is visible as a dip in the DOS at 13 eV below $E_F$
for the largest lattice constants, see Fig. \ref{fig1}.

An approximate value for the pressure, $P$, calculated as a surface integral using the virial theorem \cite{pet}, is
useful to get insight to the relative contribution from different atoms. These total pressures ($P$)
amount to 0.98 and -0.3 Mbar at the two extreme lattice constants. The partial S- and H-pressures
increases much more on H than on S when the lattice constant is reduced. This together with the charge transfers,
indicate that the S sublattice is more compressible than that of H. A charge transfer from S
to H at high pressure will enforce the hardening of H.
More precise values of $P$, shown in Table \ref{tab2}, are obtained from the volume derivative 
of the calculated total energies, and
we obtain 180 GPa for $a$=5.5 a.u. and 73 GPa for $a$=6.1 a.u., which agree well with
ref. \cite{papa}.

 \begin{table}[ht]
\caption{\label{tab2}
Lattice constants $a$, DOS at $E_F$ (in units of states/eV/Cell), pressure $P$,
the Hopfield parameter $NI^2$ in $eV/\AA^2$ for S and H, $\lambda$,
and T$_C$ estimated from the McMillan formula. The phonon moments are 600 and 1800 K for S and
H, respectively, and $\omega_{log}$ is 1500 K.}
  \vskip 2mm
  \begin{center}
  \begin{tabular}{l c c c c c c}
  \hline
$a$ (a.u.)  & $N(E_F)$ & $P$ (GPa) & $NI^2(S)$ & $NI^2(H)$ & $\lambda$ & T$_C$ (K) \\
  \hline \hline
5.4    & 1.00 & 300 & 5.6 & 4.3 & 1.0   & 145 \\
5.6    & 0.99 & 215 & 5.5 & 4.3 & 1.0   & 143 \\
5.8    & 0.90 & 150 & 4.5 & 3.4 & 0.81 & 109 \\
6.0    & 0.88 &  95  & 4.1 & 3.1 & 0.73 &  93 \\
6.2    & 0.90 &  50  & 3.9 & 2.8 & 0.70 &  73 \\
  \hline
  \end{tabular}
  \end{center}
  \end{table}

The electronic numerator $NI^2$ of the electron-phonon coupling constant $\lambda  = NI^2/K$
(where the force constant $K=M\omega^2$ is taken from experiment) is calculated in the Rigid Muffin-Tin Approximation
(RMTA) \cite{gg,daco}. The matrix elements for $d-f$-scattering in S and $p-d$ scattering
in H are missing, which leads to some underestimation of the total $\lambda$. Note also that LMTO
uses overlapping Wigner-Seitz (WS) spheres with no contribution to $I^2$ from the interstitial region. The values of $NI^2$ on H
and S are reasonably large in comparison to the values in transition metals. The low local DOS
is compensated by larger $I^2$ and the scattering to p-states, while in the A15's large scattering
arises mainly from d-states.
Our $N(E_F)$, shown in Table \ref{tab2}, are in agreement with the work of Papaconstantopoulos et al. \cite{papa},
which quite unusually are largest at low volume and large $P$. This is because of the band edges that cross $E_F$ and become
occupied at high $P$. But for even larger volume $N(E_F)$ behaves normally again and it increases when $P$ goes down, as shown by Papaconstantopoulos el al. \cite{papa}.
The matrix elements $I^2$ show strong $P$-dependences, and $NI^2$ (the so-called Hopfield parameter) increases steadily
with $P$ in agreement with ref. \cite{papa}. The absolute value of $\lambda$ is smaller than in ref. \cite{papa}.
This can partly be due to the smaller basis in our case, and also because of the use of WS-spheres in LMTO instead of non-overlapping
MT-sphere geometry in LAPW \cite{papa}.

In the estimation of  $T_c$ from the McMillan equation another uncertainty concerns the Coulomb repulsion $\mu^*$. 
Many $T_C$-calculations use $\mu^*$=0.11-0.13,
but theories for calculation of $\mu^*$ are approximate or unreliable \cite{bauer}. Retardation can make $\mu^*$ large
and screening makes it small or even
negative depending on the band width and the phonon frequency \cite{bauer}. Here we use an empirical formula for $\mu^*$ proposed by
Bennemann-Garland \cite{benne}, which leads to a small value of the order 0.03. Thus, we caution that even if our calculated
 $T_c$ from the McMillan formula will be large it is very approximate, similar to what has been concluded in the other works.  
Our $T_c$'s are of the order 145 and 75 K
between the two extreme lattice constants when the total coupling strength goes from 1.0 and 0.7.
These $T_c$'s are reduced to 95 and 40 K between the extreme values of $a$ if $\mu^*$ is 0.13. 
The different contributions to the total $\lambda$ from each S and H atoms
are comparable. Moreover, the total $\lambda$ itself is not unusually large. Using the McMillan formula, valid only
for a single effective band BCS system in the dirty limit, the high $T_c$
 is mainly because of the large phonon frequency pre-factor ($\omega_{log}$) in the equation. 

In order to investigate the pressure effects on the electronic structure 
we show in  Fig. \ref{fig1} the different total DOS for different lattice constant $a$ of the perovskite structure changing from 6.2 to  5.6 a.u.
The total DOS at $E_F$ per S-atom is about 3.5 $(Ry)^{-1}$
and 1 $(Ry)^{-1}$ per H-site, compared to the order 20 per V in V$_3$Si or in elementary V and Nb.
This is not surprising because of the large width of the wide band in H$_3$S having its bottom at 2 Ry below the chemical potential, see Fig. \ref{fig1}, while for increasing number of (d-) electrons in transition metal
A15 compounds the total band width is more like 2/3 of a Ry \cite{jmp}.

The charge within the H WS-sphere increases from 1.3 to 1.4 el./H when the
lattice constant decreases from 6.2 to 5.6 a.u., and the H-s charge goes from 0.95 to 1.0.
This fact justifies somewhat the use of LSDA for H even though atomic H with exactly 1 electron is
best described by the Hartree potential only.
The results show a very large effect of the pressure on the lowest dispersive bands with H s character, 
and the energy shift of the narrow peak of DOS at the chemical potential due to the van Hove singularity.
The shift is small in the figure because of the large energy scale.
\\
\\ 
\maketitle
{\bf Hydrogen Zero Point Motion Effects}
\\

Usually the atomic velocities ($v_i$) from the vibrations
are much slower than the electronic velocities, and the electronic structure can relax adiabatically at all times.
Therefore, the electronic structure calculations usually can neglect $v_i$, as in 'frozen-phonon' calculations. 
However it is well-known that the lattice of real materials becomes distorted at large $T$ ('thermal disorder') and
 some distortion remains at $T$=0 due to the 'zero point motion', ZPM.
 Thermal disorder and ZPM will modify and broaden the bands compared to the case of a perfect lattice 
because the potentials at different sites are not exactly the same.
Each atomic (i) lattice position deviates from its average position by $u(t,T)$, because of the excitation of phonons
The time average of $u(T)$ for harmonic uncorrelated vibrations are well represented by a Gaussian
distribution function with width (FWHM) $<u>$, which tends to $\frac{3}{2} \hbar \omega / K$ at low $T$ (ZPM)
and to $3 k_BT/K$ at high $T$, see ref. \cite{grim}. 

The Debye temperature for H phonons ($\sim 1500 K$) in sulfur hydride is much higher than $T_c$ 
and the disorder amplitude from ZPM is almost the same as at $T = T_c$.
The maximum superconducting gap (2$\Delta$) would be $\sim$ 65 meV at $T$=0 in the large Fermi surface, 
i.e., the band No.2 using the notations to identify the 5 bands crossing the Fermi level introduced in ref. \cite{bia1}. 
With the parameters as in the calculation of $\lambda$, and $\Delta = V$, the amplitude for the potential
modulation for phonons \cite{ssc}, one can estimate that $u_{\Delta}$, the displacement of
phonons that lead to superconductivity should be $\sim$0.02 \AA, i.e., less than
the amplitude of ZPM for H, or comparable to the ZPM of S.

We have used a first-principle DFT approach based on a large supercell to calculate the effect of ZPM on the van Hove singularity. 
This method it is more suitable 
for complicated 3-D materials such as $H_3$S (having a unit cell of 64 atoms) compared with alternative perturbative 
approaches for evaluation of the band energy changes 
as function of displacement (u) based on the Allen-Heine-Cardona (AHC) method, which is suitable for simple systems like carbon nanotubes \cite{capaz,gomez}.
While AHC type approach may seem more sophisticated, when it is applied to complex 3-D materials it is more
important to consider good statistics and rely on experimental information on force constants \cite{fesi}-\cite{eri14}. 
The important point of our approach is that a large supercell is used in order to have a good statistics for the individual u 
of each atom so that the bands energy changes as function of increasing disorder  can be determined properly. 
The average atomic displacements and their 3-D distribution are determined from the phonon spectrum and 
atomic masses and it does not make sense to attempt an ab-initio calculation of u using frozen 
zone-boundary phonons, and tight-binding  method. 

 Calculations for disorder within supercells with 64 atoms for FeSi \cite{fesi}, or even less (48 atoms) for purple bronze \cite{bron}, have
 shown that different generations of internal disorder in the cells produce similar results (as long as the disorder amplitude
 $<u>$ are the same).  Symmetry makes the bands identical in different irreducible
 Brillouin zones (IBZ) for ordered supercells, and these bands have their exact correspondence in the bands of the small
 cell. But this
 symmetry is lost if the atoms are disordered, and the bands have to be determined in half of the BZ.

 \begin{figure}
\includegraphics[height=8.0cm,width=8.0cm]{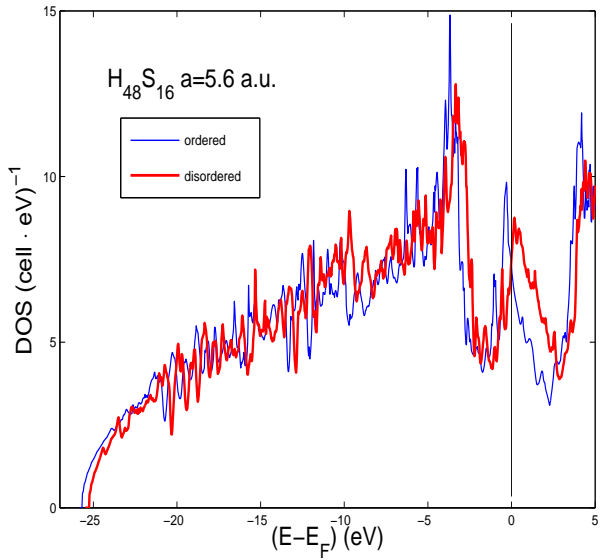}

\caption{(Color online)
 The DOS for H$_3$S for cubic 64-site supercells.
The (blue) thin line is the DOS for the perfectly ordered supercell. The (red) heavy
line shows the DOS for the disordered supercell with $u_S$=0.01a and $u_H$=0.033a
 Zero Point Motion  }

\label{fig1a}
\end{figure}

With a force constant
$K = M \omega^2$ of 7 eV/\AA$^2$ we obtain an average amplitude $<u>$ of the order 0.15 \AA for ZPM. This
is close to 10 percent of the H-H distance, which according to the Lindemann criterion suggests
that the H sublattice is near melting \cite{grim}. The $<u>$-amplitude for the S sublattice is normal, because of its
large mass, and it is probable that the rigid S-lattice is important for the stability of the structure in which
the H-atoms are rather loosely attached to their ideal positions.

The ZPM generates different shifts of the Madelung potential at different sites. This
modifies the band energies ($\Delta\epsilon$) and leads to 'fluctuations' (there is a spread of the eigenvalues in different IBZ) 
of $\Delta\epsilon$, in particular at 
the band edges. The low-T energy band fluctuations in materials with narrower band widths are known to
be about 20 meV for $u$ in the range 0.03-0.04 \AA \cite{fesi}-\cite{eri14}.
From an extrapolation of these values to the conditions in H$_3$S (larger $<u>$)
we estimate that the band energy fluctuation can be larger than 150 meV for H-bands.

The band No.2 of H$_3$S in Fig. \ref{fig1} is very wide, about 2 Ry for the high lying valence bands, 
and the S-s band overlaps with the S-p band. 
This makes the band dispersion and Fermi velocities high in large portions of the k space,
where the effects of energy band broadening in these points of the k-space is less important. 
However in portions of the K-space around the $\Gamma$-X-M path in the sc BZ, where the van-Hove 
singularity crosses the chemical potential the Fermi velocity is getting small and strong dynamical
 fluctuations controlled by the zero point lattice fluctuations are expected to be relevant.
 
The electronic calculations for the ZPM in H$_3$S  at P=210 GPa,  $a=$5.6 a.u., have been carried out using a large supercell  in which the lattice
is disordered. Each atom is assigned randomized displacements, $u_x,u_y,u_z$ in such a way that the distribution
of all displacement amplitudes ($\mid u \mid$) has a bell shaped (Gaussian) distribution with the FWHM
width equal to the averaged displacement amplitude ($<u>$) at was described before and in ref. \cite{fesi}.
In Fig. \ref{fig1a}  we show the energy  renormalization of the DOS due to the calculated zero point motion (ZPM) .
Here we consider a
2x2x2 extension of the cubic unit cell with 64 atoms totally which permits to calculate
$u$ from 192 displacement vectors, which is a reasonably good statistics for calculating $<u>$.
As was discussed above, it is sufficient to do the calculation for one disordered configuration when the supercells are
large with at least 48 atoms.
Because of the large mass difference between S and H we here allow larger $<u>$ for H ($u_H$) than for S
($u_S$). For small $<u>$ it can be shown that correlation of vibrational movements and anharmonic
terms are small \cite{grim}. But $u_H$ is large, and in the generation of
a disordered configuration for $u_H/a$=0.05, as for the expected ZPM, several pairs of H come too close to
each other. Therefore, in order to avoid large anharmonic effects at this stage,
we calculate the electronic structure for a supercell with $u_S/a$=0.01
(which is close to the expected ZPM for S) and $u_H/a$=0.033 (which is 2/3 of the expected ZPM for H).

The resulting DOS for the disordered lattice is shown in Fig. \ref{fig1a}. The 
results show a strong effect of ZPM on the energy of the narrow peak of the DOS
just below the chemical potential. This peak
has a large fraction of H s- and p-states, while the peak further down (3-4 eV below $E_F$)
has less H-character and is less affected by disorder. The width of the DOS peak with large character of H-states
gets wider in the disordered case, which is expected, but we see also a large energy renormalization of this narrow peak.
This peak which was mostly below the chemical potential in the ordered lattice is now pushed up above the chemical potential 
because of the lattice ZPM. The shift at the top of the peak is of the order of 600 meV (see Fig. \ref{fig1a}) 
 and the peak becomes broader in the disordered case.  
 
Our  simple approach to treat the ZPM has been based on the assumption that the frozen disorder in our supercell
calculation is representative for ZPM. If so we may ask how much ZPM would change in the RMTA value
of $\lambda$ and the McMillan estimate of $T_c$. The total DOS at $E_F$ is almost the same
(7.3 and 7.4 $states/eV/cell$ for ordered and disordered case, respectively) as can be seen in Fig. \ref{fig1a}.
The $\lambda's$ and $T_c's$ are calculated to be 0.88 and 0.86, and 122$K$ and 118$K$, for ordered and disordered
cases, respectively. Thus there is a small reduction of $\lambda$ and $T_c$ from disorder
even if the two $N(E_F)$ are comparable. However, one can note that the peak in the DOS has moved from being
below the chemical potential in the ordered case to be above when ZPM is taken into account indicating a 600 meV energy shift 
of the van Hove singularity with zero point lattice fluctuations involving H atoms.
In the remaining sections we discuss some details about the Lifshitz transitions as a function of pressure.
\\
\\ 
\maketitle
{\bf Lifshitz Transitions as a function of pressure}
\\

The energy shift of the van Hove singularity (vHz) can be followed by looking at the shift of the narrow peak in the total DOS near the chemical
 potential at different lattice parameters $a$ which is shown 
in Fig. \ref{fig2}.

\begin{figure}
\includegraphics[height=8.0cm,width=7.0cm]{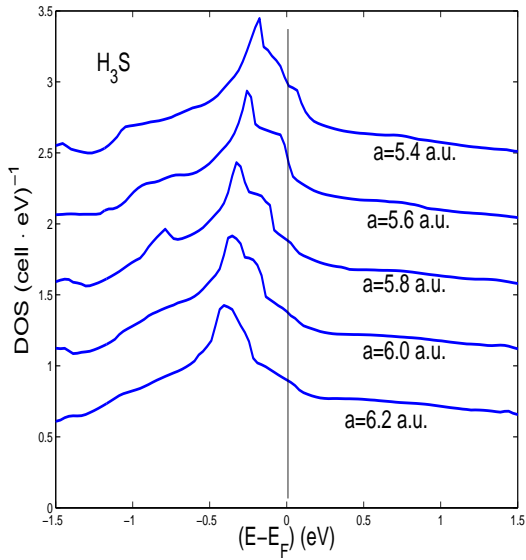}
\caption{(Color online)  The total DOS functions in Fig. \ref{fig1} on a finer energy scale near the chemical potential 
as a function of the lattice constant.  Each DOS curve is
separated by 0.5 $(cell ~ eV)^{-1}$ units for the sake of visibility.}
\label{fig2}
\end{figure}

\begin{figure}
\includegraphics[height=6.0cm,width=7cm]{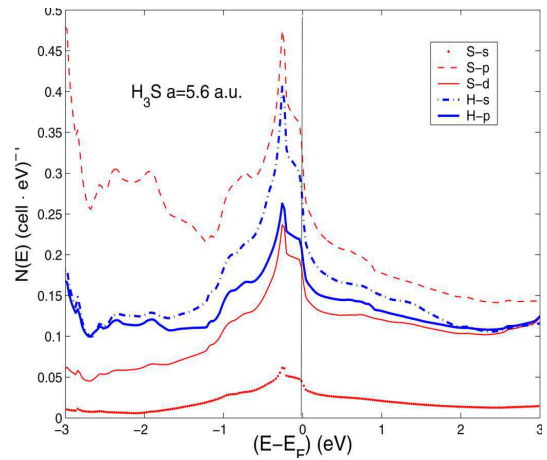}
\caption{(Color online) The partial DOS with different atomic and orbital symmetry for 
H$_3$S at the lattice constant 5.4 a.u..,
the (red) thin lines are for S and the (blue) heavy lines for H.}
\label{figpart}
\end{figure}

\begin{figure}
\includegraphics[height=6.0cm,width=7.0cm]{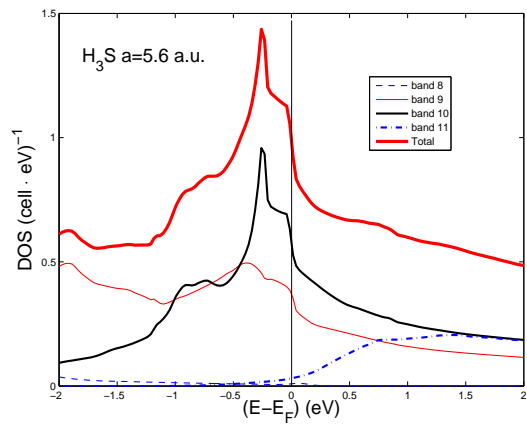}
\caption{(Color online) The partial DOS for H$_3$S at the fixed lattice constant 5.4 a.u.  
from different bands crossing the chemical potential in the simple cubic Brillouin Zone.}
\label{figpartB}
\end{figure}

\begin{figure}
\includegraphics[height=9.0cm,width=7cm]{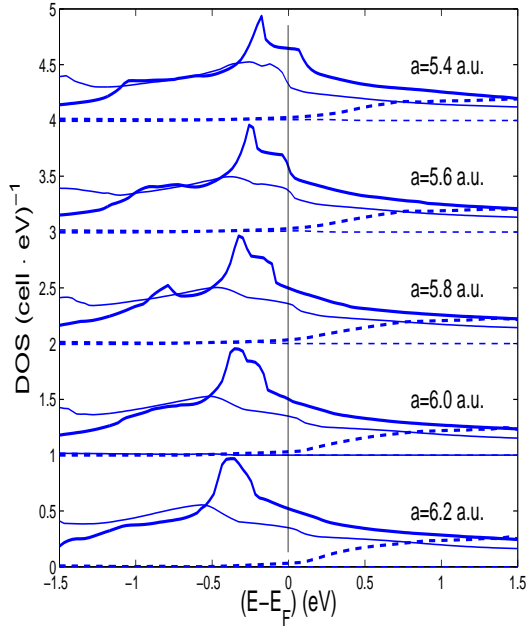}
\caption{The evolution of the partial DOS from band 8-11 in the simple cubic Brillouin Zone as function of the lattice constant.
Only bands 9 and 10, shown by thin and heavy continuos lines respectively, contribute much to the total
DOS at the chemical potential $E_F$.}
\label{fig2p}
\end{figure}

This peak is mostly due sulfur orbital contributions as shown by the partial DOS functions of the vHs for 
the case of lattice parameter  $a=5.6 a.u.$  shown in  Fig. \ref{figpart}.
The S-p orbital and S-d orbital contribute to about 30 and 16 percent of the total DOS respectively, therefore the vHs is mainly due sulfur atoms.
Both the total charge and the $\ell$-character at $E_F$ on H are mostly $s$
(60-65 percent),
and the hybridization between the s-electron and states on the S atoms away from the chains is large.
This large on-site hybridization is favorable for large dipole matrix-element contributions to the
electron-phonon coupling.
Fig. \ref{figpartB} shows the band decomposition of the total DOS in the cubic 8-site unit cell in the sc BZ into 4 bands classified as No.8, No 9, No. 10, No. 11.
The vHs and the peak in the DOS is due only to band 10 which gives the largest Fermi surface and it is
due to the flat dispersing bands with low Fermi velocity near X and M points giving a van Hove singularity.
Fig. \ref{fig2p}  shows the pressure dependence of the vHs in band 10 which goes through the chemical potential 
at about 210 GPa where a is about 5.6 a.u.. 
This result shows that the  van Hove singularity approaches 
the chemical potential at 210 GPa but it remains near the chemical potential 
in the energy range of the energy cut off of the pairing interaction in the pressure range showing high temperature superconductivity. 
Where the vHs crosses the chemical potential a Neck-Disrupting  Lifshitz Transition, of type 2,  occurs, 
Here the topology of the large Fermi surface changes because of the appearing of small pieces of tubular 2D Fermi surfaces connecting the large
 petals as discussed in the reference \cite{bia1}.
 In these tubular portions the Fermi velocity is small therefore the Migdal approximation breaks down. 
On the contrary in the large petals the Fermi energy is much larger and the Migdal approximation is valid. 

\begin{figure}
\includegraphics[height=7.0cm,width=8.0cm]{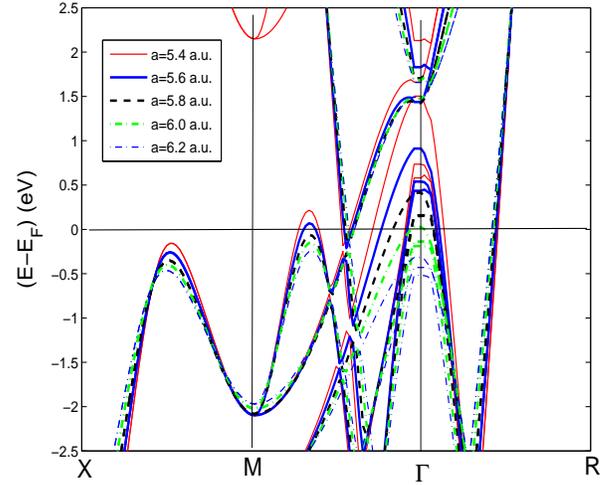}
\caption{(Color online) The band structure for H$_3$S along $X-M-\Gamma-R$ for $a$ for 
the simple cubic double cell (sc BZ) with the lattice parameter changing between 5.4
and 6.2 a.u.}
\label{7figXMGR}
\end{figure}

\begin{figure}
\includegraphics[height=7.0cm,width=8.0cm]{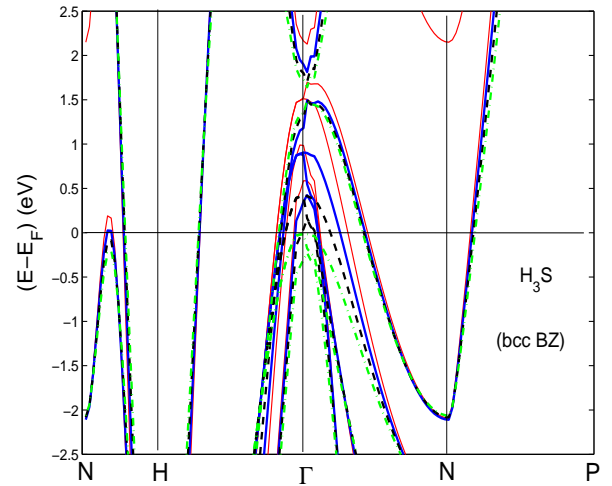}
\caption{(Color online) The variation of the band structure near the chemical potential plotted for 
the  small bcc unit cell of one formula unit of H$_3$S (bcc BZ) as a function of the lattice constants between 5.4 and 6.2 a.u..
The colors of the bands correspond with different lattice parameters as in Fig. 9.
The labels of symmetry points are as for bcc BZ. }
\label{fig9}
\end{figure}

\begin{figure}
\includegraphics[height=12.0cm,width=7.0cm]{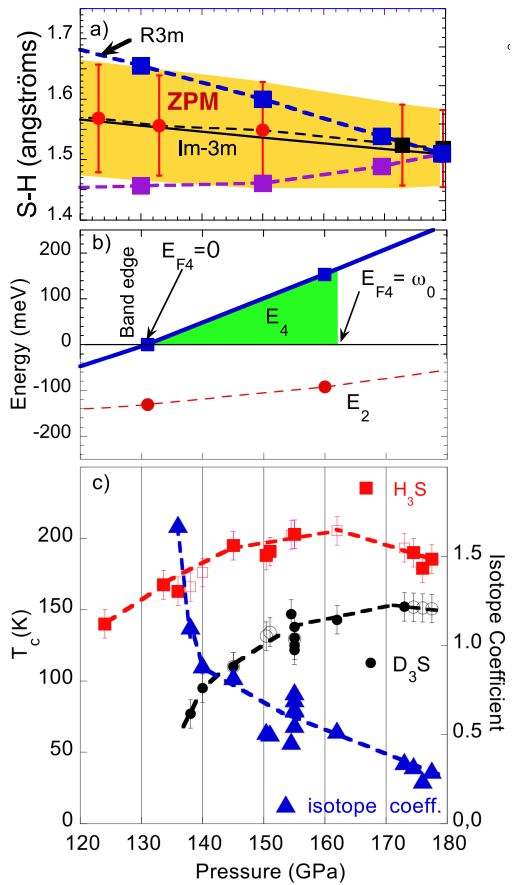}
\caption{(Color online) 
The upper panel $a$ shows the S-H bond length in the pressure range between 120 and 180 GPa where without the hydrogen ZPM, the R3m structure was expected to be stable with the S-H bond splitting into a long (blue squares) and short (violet squares) sulfur-hydrogen bond. The amplitude of the calculated ZPM of the S-H bond is indicated by the red error bars \cite{bia1,bia2}. The S-H amplitude of the zero point motion, ZPM, is larger than the S-H splitting in the range 130-180 GPa therefore in this pressure range the ZPM stabilizes the $Im\bar{3}m$ structure \cite{bia2} in agreement with experiments \cite{ere0}. 
The panel $b$ shows the Fermi energy E$_F$$_4$ in the small hole Fermi
 pocket at $\Gamma$  as a function pressure. The top of this band crosses 
 the chemical potential at 130 GPa giving the Lifshitz transition of type 1 for the appearing of a new Fermi surface. 
 The position of the vHs  E$_2$  remains below the chemical potential but it remains in the energy range of pairing interaction.
Panel $c$ shows the variation
of the experimental  isotope coefficient calculated from data in ref. \cite{ere0} which shows a divergence from 0.3 at 180 GPa to 1.5 at 135 GPa,
which is not predicted by the BCS theory. The critical temperature decreases toward zero with a decrease of about 60 K in a range of 30 GPa, beteen 160 GPa and 130 GPa 
which is not predicted by the BCS theory. Both phenomena are predicted by the general theory of multigap superconductivity 
near a Lifshitz transition \cite{bia1,bia2}.}
\label{fig11}
\end{figure}

In order to associate the crossing of the chemical potential of the 
narrow peak in the DOS with the Lifshitz transitions on the topology of the Fermi surfaces 
we have plotted the electronic bands in a narrow energy range 
near the chemical potential for both the sc BZ and bcc BZ in Fig. \ref{7figXMGR}  
and Fig. \ref{fig9}.
These band plots show that several Lifshitz transitions appear for increasing pressure.
From Fig. \ref{7figXMGR} it is seen
that a local band maximum crosses $E_F$ in the sc BZ at
about 2/3 of the $\Gamma-M$ distance when $a$ is $\sim$ 5.8. a.u. (see  Figs. \ref{fig3bcc} and Figs. \ref{fig4bcc} )
and the same band crosses the chemical  potential between $N$ and $H$  in Fig. \ref{fig9}
in the bands for the bcc BZ. This Lifshitz transition occurs at $P=210 GPa$. 
The energy difference $E_2$ between the chemical potential and this local band maximum which is associated with the vHs 
goes from -200  to +100 meV when $a$ decreases from 6.2 to 5.6 a.u.. 
This gives a neck disrupting Lifshitz transition in the Fermi surface at $210 GPa$ where the neck disappears at low pressure in the N-H direction \cite{sanna}.
The Fermi surface neck appears at the point where the narrow peaks in the DOS crosses the chemical potential.
Another band is seen to be approaching the chemical potential in Fig. \ref{7figXMGR} between $X$ and $M$ when the pressure
goes up. However, this potential band crossing (which is not seen on a symmetry line
in the band plots for the small bcc cell) will not reach $E_F$ to make a FS pocket unless P is increased even more.

In the pressure range $120<P<160$ GPa, where the onset of superconductivity occurs followed by the rapid increase of the critical temperature 
there are Lifshitz transitions, of type 1,  for the appearing of new Fermi surface spots at 
$\Gamma$. 
There are 3 bands pushed up by increasing pressure which cross 
the chemical potential and three small closed Fermi surfaces appear.
In fact Fig. \ref{7figXMGR}. Fig. \ref{fig9},
show that the tops of these bands at $\Gamma$ are all above the chemical potential
at the highest $P$ while the tops of these 3 bands bands are below $E_F$ at low $P$.

In Fig. 11 we have plotted in panel $a$ the S-H bond length as a function of pressure 
and the amplitude of the calculated spread of this bond length due to the hydrogen zero point motion indicated by the red error bars. 
The USPEX theory  \cite{duan,duan2} predicts the second order phase transition from $Im\bar{3}m$ to R3m structure at 180 GPa where the sulfur atoms remain
 in the same sites of the bcc unit cell while hydrogen ions are frozen in one of the two minima of their double well potential due to the hydrogen
  bond like in ice structural transitions. 
  On the contrary this transition is forbidden in the pressure range 120-180 GPa since the quantum zero point motion amplitude
  which  is larger than the difference  between the short and long hydrogen bonds expected in the R3m structure,
  Therefore the ZPM stabilizes the $Im\bar{3}m$ structure in the pressure range 130-180 GPa  \cite{bia1} in agreement with recent experiments \cite{ere0}.
  
The panel $b$ in Fig. 11 shows the Fermi energy in the small hole Fermi pocket at $\Gamma$ as a function pressure. 
The top of this band crosses the chemical potential for a pressure larger than 130 GPa. At this pressure the Lifshitz transition of  
type 1 for the appearing of a new Fermi surface occurs. In fact above 130 GPa a new small Fermi pocket appears at the $\Gamma$  point of the Brillouin zone.
The Fermi energy E$_F$$_4$  remains smaller than the energy cut off of the pairing interaction below 160 GPa. Therefore In the pressure range between 
130 and 160 GPa the Migdal approximation breaks down for electron 
pairing in the small Fermi pocket at the $\Gamma$  point of the Brillouin Zone. 

Finally panel $c$ of Fig, 11 shows the variation
of the experimental  isotope coefficient in this pressure range calculated  
by recent data \cite{ere0} which shows a regular increase with decreasing pressure from 0.3 at 180 GPa to 1.5 at 135 GPa. 
The divergence of the isotope coefficient 
approaching the Lifshitz transition at 130 GPa is not predicted by the BCS theory
using standard Midgal approximation and an effective single band but it is predicted 
in the frame of general theory of multigap superconductiivity near a Lifshitz transition \cite{bia2}
This is supported by the decrease of the critical temperature of about 40 K in $H_3$S and of about  60 K in $D_3$S  while in the BCS  calculations 
the variation of the critical temperature over this 30 GPa range is predicted to be of the order of 10 K. 
In fact in the multigaps superconductors at a Lifshitz transition  the critical temperature goes toward zero (like at the Fano-Feshbach anti-resonance) at the appearing of a new nth Fermi surface where electrons in the nth band have zero energy  $E_F$$n$ =0 at the band edge forming a BEC condensate, while the critical temperature has a maximum (Fano-Feshbach resonance) where the electrons have a Fermi energy of the order of the pairing energy forming a condensate in the BEC-BCS crossover \cite{bia2}.
\\
\\ 
\maketitle
{\bf Conclusions}
\\

In $H_3$S the onset and the maximum superconducting critical temperature, 203 K, are controlled by pressure, like in cuprates where the onset and the
maximum value of the critical temperature 160 K is reached by tuning the lattice misfit strain at fixed doping \cite{agre}.

The calculated band structure for an ordered $H_3$S lattice as a function of pressure clearly shows multiple Lifshitz
transitions for appearing of new Fermi surface spots in the pressure range showing high $T_c$ superconductivity,
which together with quantum hydrogen zero point motion puts the system beyond the Migdal approximation.

New Fermi surface spots appear at the $\Gamma$  point at 130 GPa pressure where the onset of the high critical temperature  
appears. It is possible that the appearing new Fermi surface spots drive the negative interference effect in the exchange
interaction between multiple gaps \cite{shape} contributing to the suppression of
the critical  temperature \cite{bia1}. This is supported by the isotope coefficient which diverges 
at 130 GPa reaching a value of 1.5  \cite{bia1}, see Fig. \ref{fig11}, well beyond the predictions of single band Eliashberg theory.
The divergence of the isotope coefficient observed here has been already observed
in cuprates \cite{bus1,per1} providing a clear experimental indication for a
unconventional superconductivity near a Lifshitz transition \cite{bia2}. 

Increasing the pressure to 210 GPa a van Hove singularity crosses the chemical potential giving a Lifshitz transition for opening a neck.
Moreover the vhs remains near the chemical potential within the energy range of the 
energy cutoff for the pairing interaction over the full pressure range between 210 and 260 GPa.
We show that the quantum zero point hydrogen fluctuations in a double well  \cite{bia1} 
typical of hydrogen bond and involving the $T_u$$_2$ phonon stretching mode, 
has strong effect on the electronic states near the 
Fermi level. The quantum hydrogen zero point motion, induces fluctuations of the 600 meV of the energy position of the vHs. 
The zero point amplitude of the S-H stretching mode, involving the $T_u$$_2$ phonon, stabilizes the $Im\bar{3}m$ structure 
in the pressure range 130-180 GPa and induces large fluctuations of the small Fermi surface pockets at $\Gamma$ .
In conclusion we have shown the presence of large displacement amplitudes of ZPM. Single phonon waves can be disturbed
 by lattice quantum zero point disorder \cite{eri14} but superconductivity seems to resist to perturbations
from ZPM in H$_3$S. On the other hand, it is also seen that the DOS peak at $E_F$ seems to pass through $E_F$
with the large zero point motion.  Finally more work is needed to investigate the variation of the Fermi level $E_F$ in different  Fermi surfaces with different H isotopes
which change the zero point motion amplitude. 
\\
\\
\maketitle
{\bf Author contribution statement}
\\
"T. J. and  A.B. wrote the main manuscript text and prepared figures. 
All authors reviewed the manuscript and contributed equally to the work" 
\\
\\
{\bf Additional Information}
The authors declare that they have no competing financial interests.
\\
\\
{\bf How to cite this article}:  Jarlborg, T. and Bianconi, A. Breakdown of the Migdal approximation at Lifshitz
transitions with giant zero-point motion in $H_3S$ superconductor. Sci. Rep. 6, 24816; doi: 10.1038/srep24816 (2016)
\\

\end{document}